%% file: main.tex
\documentclass{article}
\usepackage[utf8]{inputenc}
\usepackage[margin=1.2in]{geometry}
\usepackage[nottoc]{tocbibind}
\usepackage[dvipsnames]{xcolor}
\usepackage{setspace}
\usepackage[colorlinks=true, linkcolor=winered, citecolor=winered, urlcolor=winered, bookmarks=true]{hyperref} 
\usepackage{eurosym}

\definecolor{winered}{rgb}{0.5,0,0}

\usepackage{natbib} 

\title{Big Data Privacy in Emerging Market Fintech and Financial Services: A Research Agenda\thanks{\footnotesize We thank Seth Garz, Zoe Kahn, Dan Cassara, and Bilal Zia for constructive feedback on this paper. We gratefully acknowledge financial support from Wells Fargo, the Institute for Business and Social Impact, the Bill and Melinda Gates Foundation, the Center for Effective Global Action, and the National Science Foundation under CAREER Grant IIS-1942702.}
}

\author{
		Joshua E. Blumenstock \\
		    \footnotesize UC Berkeley \\
		    \footnotesize jblumenstock@berkeley.edu  \\
		    \and
		Nitin Kohli \\ 
			\footnotesize UC Berkeley  \\
			\footnotesize nitin.kohli@berkeley.edu
	}
	
\vspace{-0.5cm}	
\date{August 2023}

\begin{document}

\maketitle

\begin{abstract}
\normalsize

The data revolution in low- and middle-income countries is quickly transforming how companies approach emerging markets. As mobile phones and mobile money proliferate, they generate new streams of data that enable innovation in consumer finance, credit, and insurance. Already, this new generation of products are being used by hundreds of millions of consumers, often to use financial services for the first time. However, the collection, analysis, and use of these data, particularly from economically disadvantaged populations, raises serious privacy concerns.  This white paper describes a research agenda to advance our understanding of the problem and solution space of data privacy in emerging market fintech and financial services. We highlight five priority areas for research: conducting comprehensive landscape analyses; understanding local definitions of ``data privacy'';  documenting key sources of risk, and potential technical solutions (such as differential privacy and homomorphic encryption); improving non-technical approaches to data privacy (such as policies and practices); and understanding the tradeoffs involved in deploying privacy-enhancing solutions. Taken together, we hope this research agenda will focus attention on the multi-faceted nature of privacy in emerging markets, and catalyze efforts to develop responsible and consumer-oriented approaches to data-intensive applications.

\end{abstract}

\doublespacing

\newpage

\setlength{\parindent}{2em}
\setlength{\parskip}{1em}
\renewcommand{\baselinestretch}{1.1}

\section*{Introduction}
\label{introduction}
\input{0_introduction}




\section{Priority 1: Landscape analysis}
\label{3-1_documenting}
\input{3-1_documenting}

\section{Priority 2: Defining ``privacy'' in emerging markets}
\label{3-2_current_priv_defs}
\input{3-2_current_priv_defs}

\section{Priority 3: Technical Solutions to Data Privacy}
\label{tech_solutions_whole_section}

\subsection{Background and Motivation: Privacy Threat Analysis}
\label{3-3_vulnerabilities}
\input{3-3_vulnerabilities}

\subsection{Overview of Technical Solutions}
\label{3-4_tech_solutions}
\input{3-4_tech_solutions}

\section{Priority 4: Non-technical Approaches to Data Privacy}
\label{3-5_nontech_solutions}
\input{3-5_nontech_solutions}

\section{Priority 5: Understanding the Real-World Tradeoffs Between Privacy and Other Key Priorities}
\label{3-6_privacy_values_tradeoff}
\input{3-6_privacy_values_tradeoff}

\section{Conclusion}
\label{4_conclusion}
\input{4_conclusion}

\newpage
\bibliographystyle{aer}
\bibliography{references}

\end{document}

%% file: 0_introduction.tex

Today, 95\% of the global population has mobile-phone coverage, and the number of people who own a phone is rising fast. These devices generate troves of personal data on billions of people -- including on people who live on a few dollars a day, and who historically have been less visible in national statistics and other administrative datasets. This ``data revolution'' is transforming the way that private companies approach emerging markets, and is revolutionizing different sectors including consumer finance \citep{heaton2017deep, hambly2021recent}, credit \citep{francis2017digital}, and insurance \citep{nagendra2020satellite}. 


In one particularly striking example, new sources of data -- and new computational tools for processing those data -- have transformed the consumer lending landscape in many developing countries. A suite of immensely popular new ``digital credit'' products use data from mobile phones to construct alternative credit scores, which are now used to originate loans to hundreds of millions of historically unbanked individuals \citep{francis2017digital}. The key insight behind this technology is that some people, such as those who frequently make international calls or who have more Facebook friends than others in the same area, are more likely to repay their debts. Machine learning algorithms can detect these patterns \citep{bjorkegren2020behavior} and construct credit scores for hundreds of millions of mobile phone owners who would otherwise been excluded from formal financial services due to a lack of collateral, traditional financial history, or access to a bank.

The data generated from mobile phones is also changing how governments and humanitarian organizations approach public health management \citep{oliver2020mobile}, social protection \citep{blumenstock2020machine}, and humanitarian aid \citep{aiken2022machine}. With some modifications, the same algorithms that Google, Facebook, Amazon, and other companies use to target advertisements to people online can also be used to target resources to people living in poverty. These algorithms leverage personal data from mobile-phone networks and satellite imagery to identify ``digital signatures'' of poverty \citep{blumenstock2015predicting,blumenstock_estimating_2018}. For example, in many African countries wealthier individuals tend to make more international phone calls -- as observed in mobile-phone networks -- while poorer individuals are more likely to live in houses with thatched roofs -- as observed in satellite images \citep{blumenstock2018don}.

In addition to enabling new consumer-facing applications, the responsible use of private ``big'' data can also improve the functioning of entire sectors and industries. For instance, data sharing between financial institutions can facilitate more effective approaches to fighting money laundering \citep{sangers2019secure, van2021privacy} and fraud \citep{hipgrave2013smarter, henecka2014privacy}.\footnote{See also \textit{Data Sharing for the Prevention of Fraud: Code of practice for public authorities disclosing information to a specified anti-fraud organisation under sections 68 to 72 of the Serious Crime Act 2007} presented to the UK Parliament, March 2015.} This, in turn, can reduce the costs that are borne by consumers, institutions, and society at large.


At the same time, the widespread use of personal data -- particularly from economically disadvantaged populations -- raises serious concerns about the implications for the personal privacy of consumers \citep{de2018privacy}. As mobile phones and other digital technologies become ubiquitous in everyday life, the data captured by such devices has become  particularly sensitive. After all, these data describe more than just the digital transactions of an individual; they can also describe their way of life by revealing their associations, their behaviors, their habits, and their beliefs \citep{kohli2021leveraging}.\footnote{For more examples, see Justice Sotomayor's concurrence in \textit{US v. Jones}, 565 U.S. 400, 132 S. Ct. 945, 181 L. Ed. 2d 911 (2012).}

Thus, while these new sources of data offer immense opportunities for new commercial and humanitarian applications, it is imperative to find consistent and secure ways to process and analyze the sensitive data. Unfortunately, contemporary approaches to data privacy in many low- and middle-income countries are inconsistent and often nonexistent. A wide range of technical and non-technical ``solutions'' exist, but most are not developed with emerging market consumers or companies in mind. Regulatory frameworks are patchwork and rarely enforced \citep{ademuyiwa2020assessing}. Furthermore, scholars have questioned the ethics of using such data \citep{taylor2016no}, and point out that data sharing arrangements are frequently extractive, rather than collaborative \citep{abebe2021narratives}.  

The failure to protect personal privacy -- and the high-profile examples of the harm caused by such failures \citep[cf.][]{crawford_big_2014} -- threatens to undermine financial and humanitarian applications that could otherwise positively impact society. However, this need not be the case. Appropriately leveraged, privacy-enhancing technologies, policies, and practices have the ability to simultaneously enable innovative uses of big data while providing robust and meaningful privacy protections for personal data.




This white paper describes a research agenda to advance our understanding of the problem and solution space of big data privacy in emerging market fintech and financial services, as well as related applications of big data in low- and middle-income countries (LMICs). Central to our proposal is that robust approaches to privacy will necessarily pair technical approaches alongside privacy-enhancing policies and practices.

We highlight five priority areas for research. Section~\ref{3-1_documenting} describes the need for more comprehensive landscape analysis.  Section \ref{3-2_current_priv_defs} highlights the lack of a basic understanding of what ``privacy'' means in non-Western contexts. Section \ref{tech_solutions_whole_section} is broken into two parts, with Section~\ref{3-3_vulnerabilities} highlighting what we perceive to be five key sources of risk related to data privacy, and Section \ref{3-4_tech_solutions} outlining promising avenues for technical research to attend to these risks. Section \ref{3-5_nontech_solutions} discusses non-technical approaches to data privacy, and how they can be used to complement the technical solutions described in the prior section. Section \ref{3-6_privacy_values_tradeoff} provides a deep dive into one of the most pressing areas for focused research: the tradeoff between privacy and other common objectives, such as accuracy, consumer welfare, and profitability. Section \ref{4_conclusion} concludes with discussion of this research agenda's relevance for consumers, financial institutions, and society.

%% file: 3-1_documenting.tex
It is only in the last few years that emerging markets have seen widespread use of large-scale personal data.  And while the adoption of digital credit and other products has been rapidly on the rise -- with hundreds of millions of active users across LMICs -- we unaware of any systematic and rigorous analysis describing the landscape of big data privacy in fintech and related emerging market applications.
Developing this systematic landscape is important to understand (1) how personal data is currently being used in fintech and related applications; and (2) if and how the privacy of consumers and firms is currently being protected. We therefore propose, as a foundational step, a landscape analysis to identify major areas of opportunity and concern, and to highlight important gaps in technology, investment, policy, and practice.

While such an analysis is complicated by the historical opacity of the financial services industry, current approaches to big data privacy in the tech sector are relatively better publicized. For instance, it is increasingly common for large US-based tech firms to voluntarily disclose information about their consumer privacy practices. Google \citep{erlingsson2014rappor}, Uber \citep{near2018differential, johnson2018towards}, Amazon,\footnote{See \url{https://www.amazon.science/tag/differential-privacy}} Meta,\footnote{See \url{https://privacytech.fb.com/differential-privacy/}} Microsoft,\footnote{See \url{https://blogs.microsoft.com/ai-for-business/differential-privacy/}} Apple \citep{tang2017privacy}, and LinkedIn \citep{kenthapadi2017linkedin, kenthapadi2018pripearl} all have adopted and publicized their use of differential privacy for some of their big data applications. Additionally, Google publicly states that it also protects data privacy using $k$-anonymity\footnote{See \url{https://policies.google.com/technologies/anonymization?hl=en}} and federated learning,\footnote{See \url{https://ai.googleblog.com/2017/04/federated-learning-collaborative.html}} while Meta publicly discloses that it also uses secure multiparty computation\footnote{See \url{https://privacytech.fb.com/multi-party-computation/}} and federated learning.\footnote{See \url{https://privacytech.fb.com/on-device-learning/}} 

Outside of the big tech firms that voluntarily disclose the types of technologies they use to protect individual data, not much is known about industry practices more broadly. At times, when firms poorly attend to individual privacy, a regulatory body (such as the Federal Trade Commission) steps in and reveals information surrounding the specific actions that led to privacy harms \citep{hoofnagle2016federal}. But regulatory enforcement is far less common in emerging markets \citep{ademuyiwa2020assessing}.

\paragraph{\textit{Landscape Analysis Research Questions}} In order to understand the current landscape of big data privacy in financial services and related emerging market applications, three major research questions need to be addressed.
\begin{enumerate}
    \item \textit{Use cases:} What are the key existing and upcoming products, services, and applications that require personal data in emerging markets (in financial services and related settings)? 
    \item \textit{Data access and control:} What data are being used, where do the data come from, and who controls and regulates those data? Are data commonly shared between institutions and organizations? If so, for what purposes and under what conditions?
    \item \textit{Existing protections:} What technical and non-technical approaches are currently being used to safeguard data privacy? How are those approaches typically adopted and/or adapted? And where do these approaches fall short?
\end{enumerate}

%% file: 3-2_current_priv_defs.tex
The concept of privacy has been described in a multitude of ways by legal scholars \citep{solove2005taxonomy}, philosophers \citep{nissenbaum2004privacy}, social scientists \citep{altman1977privacy}, and technical researchers \citep{dwork2006calibrating}. Conceptions of privacy differ not only between these fields but also within each of these fields. For example, legal discourse has described privacy as an individual's ``right to be let alone'' \citep{brandeis1890right}, an individual's ability to have ``control over their infomation'' \citep{westin1968privacy}, and as a ``zone of refuge'' for an individual \citep{sklansky2014too}. 

The fact that privacy can be described in multiple ways does not mean that some notions of privacy are ``wrong.'' Nor does it mean that one notion of privacy is ``more right'' than another. Rather, privacy is best understood as a \textit{pluralistic} \citep{koops2016typology}, \textit{social} \citep{post1989social}, \textit{contextual} \citep{nissenbaum2004privacy}, and \textit{contested} \citep{mulligan2016privacy} concept. It is plural in the sense that privacy can mean multiple things. It is social in that privacy is about individuals and groups of individuals. It is contextual in that actions that violate privacy in one setting may be appropriate in another.\footnote{For example, within the US, the warrantless wiretapping of an individual's house by the police has been viewed as a privacy violation, but the wiretapping of an individual's house for whom the police have secured a warrant for is not viewed as such.} And it is contested in sense that disagreements about ``what privacy is'' are essential to provide robust privacy protections across different contexts. Taken together, these imply that there is not a single definition of privacy that can be applied in every setting. 

This has two implications for \textit{data privacy}. First, this means that there is no ``silver bullet'': no single technology, law, or regulation can be used to address every single privacy issue stemming from the use of data \citep{mulligan2016privacy}. Second, this means that the design of data privacy protections should be aimed at addressing privacy \textit{for a specific setting}. Data collected from different contexts have different sensitivities, and thus require different notions of privacy; data privacy protections should be calibrated based on the particular setting in which they are being deployed.

Much of our current understanding of privacy is rooted in a Western perspective. For example, the above notion of privacy as a ``right to be let alone'' first appeared in a Harvard Law Review article in the 1890's as a reaction to the rise of photography and printed media to capture and distribute images of individuals' private lives \citep{brandeis1890right}. Similarly, the notion of of privacy as ``control over information'' arose in the US in the 1970's. Fueled by concerns surrounding the adoption and expansion of government record-keeping systems \citep{ware1973records}, as well as inappropriate government surveillance activities, including the Watergate scandal and the wiretapping of Martin Luther King Jr \citep{bajpai2017privacy}, the notion of privacy as ``control over information'' sought to mitigate the harms posed by secretive surveillance systems. In addition to being US-centric, these notions of privacy were articulated to deal with the specific social problems of their time.

However, the most salient privacy concerns in emerging markets are not necessarily the same as those that are most important in wealthy nations. For instance, \citet{abebe2021narratives} have  documented that some African communities view the sharing of aggregate statistics about their community with individuals outside their community as a privacy violation, regardless of the technical protections used to secure the information. This conception differs from the US, where the public disclosure of statistics is common and at times required by law \citep{abowd2018us}.

Differences in privacy concerns are not exclusively confined to data sharing practices. Rather, these differences can permeate throughout everyday interactions. For example, within the US it is common to present ID when purchasing over-the-counter medication at a pharmacy. However, in Saudi Arabia this practice is not common except in special circumstances, so the requirement to disclose your identity to purchase over-the-counter medication can be viewed as an intrusive solicitation for too many details about the consumer \citep{abokhodair2016privacy}, resulting in a privacy violation. More generally, on-the-ground concerns about privacy in Lagos and S$\tilde{a}$o Paolo may fundamentally diverge from those found on the ground in Los Angeles and San Francisco.

\paragraph{\textit{Privacy Definition Research Questions}} Given that much of our understanding of data privacy -- and privacy more generally -- is rooted in the US context, there is a need to document how privacy is conceptualized \textit{locally} by individuals whose privacy is at risk in these emerging markets. A nuanced and comprehensive understanding of data privacy concerns is a necessary precursor to developing technical and non-technical solutions to address those concerns. Failing to do this work risks developing solutions that do not actually address the concerns of people in their everyday lives. Examples of outstanding questions include:

\begin{enumerate}
    \item \textit{Definitions:} How do specific communities and constituents conceptualize data privacy? How do consumers conceptualize the privacy harms that could arise from the use of personal data sources for financial decisions? How much do these definitions and concerns vary from one region to another? What commonalities exist?
    \item \textit{Methodology,  Conceptualization, and Measurement:} How should researchers engage local communities in conversations surrounding privacy in a meaningful way? Once articulated, how can privacy concerns best be captured and characterized? To what extent do Silicon Valley approaches to quantifying privacy (or the loss thereof) adequately address the local concerns, norms, and expectations of privacy in emerging market contexts?
    \item \textit{Tradeoffs:} When centered in local perspectives, what is the relative value of data privacy, vis-à-vis other common tradeoffs (e.g., product innovation, availability, and cost) in different local contexts?
\end{enumerate}

%% file: 3-3_vulnerabilities.tex
 Data privacy is typically described in relation to privacy harms that arise from the (i) collection, (ii) storage, (iii) processing, or (iv) dissemination of information \citep{solove2005taxonomy}. While this 4-part categorization points to \textit{when} failures of data privacy can occur when data are used, it provides an incomplete view of privacy harms. For example, it does not tell us \textit{who} infringed upon privacy, \textit{what} this infringement looks like from a technical perspective and for the individual, \textit{where} the vulnerability and threat came from, and \textit{why} and \textit{how} it occurred. In other words, this 4-part categorization under-specifies the ``who-what-when-where-why-hows'' of actions that violated privacy.

Since, as discussed in Section~\ref{3-2_current_priv_defs}, privacy is inherently a social concept, the analysis of privacy should center around the individuals and groups whose privacy is at stake. This motivates a clear understanding of \textit{how} individuals can be harmed by their data. Broadly, we view data privacy harms (that occur after data are collected\footnote{Our focus on data privacy harms that occur post-collection does not imply that the collection of data can never constitute a privacy violation. Rather, we focus on post-collection data privacy harms in this section since we will be documenting five different ways in which harmful information can be inferred from personal data that has already been collected.}) as a two-step process: data \textit{reveals something} about an individual, and from these revelations \textit{something occurs} that negatively impacts an individual. Examples of the latter include the disclosure of information (such as doxing) or the threat of disclosure of information (such as blackmail), financial harm, and physical harm \citep{solove2005taxonomy}, as well as reputational harms and changes in the opinion of the individual who information was revealed.

How can potentially harmful information be revealed about individuals from their data? We organize our discussion according to five different types of revelations: (1) revelations from one's own data; (2) revelations from another individual's data; (3) revelations from the social meaning of the dataset; (4) re-identification of individuals from ``anonymized'' data; and (5) reconstruction of data from statistics. We explain each of these 5 vulnerabilities in turn below.

\subsubsection{Privacy threats from an individual's own data}
\label{priv_threat_own}

An individual's data can reveal much more about them than first meets the eye. As a motivating example, consider a dataset that contains the locations an individual visited over the past month. Such datasets can be useful for a wide range of tasks, ranging from city planning purposes \citep{bassolas2019hierarchical} to humanitarian response \citep{kohli2023privacy}. 

While these locations may ``only'' represent geographical coordinates on a map, the social meaning attributed to the physical locations can be used to draw inferences about an individual's life, including their sexual preferences \citep{ovide2021, boorstein2021} and political preferences \citep{thompson2019twelve}. For example, a visit to place of worship evokes different social perceptions about the individual than a visit to a gentleman's club, which evokes different perceptions than a visit to a drug or alcohol rehabilitation center. 

The process of generating these inferences by leveraging the social meaning of data are called \textit{social meaning attacks}, and they can be notoriously hard to defend against. When multiple datapoints are collected about an individual, social meaning attacks can paint an even more detailed picture about an individual's life \citep{ohm2009broken}. For instance, a visit to an OB/GYN followed a week later by a visit to an expecting mother's clothing store describes more about an individual's life than a visit to each of these locations in isolation.\footnote{This is inspired by an example provided in \textit{US v. Maynard}, 615 F.3d 544 (D.C. Cir. 2010), where the DC Circuit Court remarked that the ``sequence of a person's movements may reveal more than the individual movements of which it is composed.''}


\noindent \textit{Takeaway:} While the literal representation of data may seem innocuous, the information that the data reveals can be far more noxious for individual privacy, particularly when combined with the social meaning of data. 

\subsubsection{Privacy threats from other people's data}

Inferences based on similarly situated individuals can also lead to revelations that are noxious for individual privacy. This type of \textit{predictive privacy harm} \citep{crawford2014big} arises in situations where statistical associations reveal private information that an individual has not disclosed. 

Examples of this abound. In 2009, researchers showed that it was possible to infer a Facebook user's sexual orientation based on their social network \citep{jernigan2009gaydar}. For individuals who do not choose to disclose their sexual orientation on Facebook, the use of predictive technologies to infer this sensitive fact creates unique privacy challenges. And in 2012, Target discovered that a teenage girl was pregnant based on the similarity between her shopping patterns and those of other pregnant women, despite her never revealing to Target that she was pregnant \citep{hill2012target}. 

\noindent \textit{Takeaway:} Big data technologies can leverage statistical associations to infer non-disclosed attributes, and reveal sensitive information that individuals may have intentionally withheld. 

\subsubsection{Privacy threats from the social meaning of a dataset}

In Section \ref{priv_threat_own}, we described how social meaning attacks can reveal information about an individual beyond what is explicitly stated in their data. One specific type of social meaning attack occurs when the ``type of database'' an individual is associated with reveals information about them. 

One such example is the Ashley Madison data breach. For context, Ashley Madison is an online dating site for married individuals who are looking to have an affair. In 2015, hackers published troves of the company's data online, including their members' information \citep{mansfield2015ashley}. This included each member's name, email address, relationship status, gender, birthdate, billing address, and debit / credit card information.\footnote{See Paragraph 37 in the Federal Trade Commission's complaint against Ashley Madison, available \href{https://www.ftc.gov/system/files/documents/cases/161214ashleymadisoncmplt1.pdf}{here}.} This online repository of user details would eventually come to be known as the infamous Ashley Madison dataset. 

While the release of this hacked information constitutes a privacy breach in its own right, the social meaning of the Ashley Madison dataset automatically reveals even more salacious information about each individual in the dataset. Since Ashley Madison was a website for married individuals seeking an extramarital affair, the existence of a user profile in the Ashley Madison dataset implies that the individual was seeking (or had) an affair. That is, the social meaning of the dataset reveals a characteristic about the individuals whose data is housed within the dataset. 

This is not an isolated incident. \cite{homer2008resolving} were able to determine the involvement of an individual in a Genome Wide Association Study based on aggregate statistics.\footnote{The technical process by which this involvement was determined is example of a \textit{membership inference attack} \citep{shokri2017membership}, also known as a \textit{tracing attack} \citep{dwork2017exposed}. We will discuss these attacks in more detail in Section \ref{priv_threats_recon_mem}.} Since all the individuals in this study shared a common disease, the discovery of an individual in this dataset reveals that the individual has the disease as well.

\noindent \textit{Takeaway:} The specific values of an individual's data record are not required to reveal a sensitive fact about them. For datasets with social stigma or particular sensitivities about them, simply being identified as part of such a dataset can reveal sensitive information about an individual. 

\subsubsection{Privacy threats from ``anonymized'' datasets}

When data are used for computational purposes, they are typically stripped of overtly identifying information, such as an individual's name, social security number, phone number, etc. Due to the lack of an explicit identifier, such data are often assumed to be ``anonymized.'' 

However, the absence of an identifier within a dataset does not preclude the remaining information in the dataset from de facto identifying an individual \citep{narayanan2014no, kroll2019privacy}. For example, in the late 1990's, Latanya Sweeney was able to re-identify the then governor of Massachusetts, William Weld, in a supposedly ``anonymized'' medical dataset by linking his birthdate, zip code, and sex to a voter registration database that contained his name \citep{sweeney2002k}. Similar methods have been successfully re-identified individuals in ``anonymized'' datasets in different domains, including Netflix data \citep{narayanan2008robust}, HIPAA-compliant hospital data \citep{yoo2018risks}, and clinical drug study data \citep{branson2020evaluating}. For location data in particular,  \cite{de2018privacy} famously showed that four spatio-temporal datapoints were sufficient to uniquely identify 95\% of individuals in an ``anonymized'' mobile phone dataset.

The method of linking fields in an ``anonymized'' record to another data source is known as a \textit{linkage attack}, and is an example of a \textit{re-identification attack} more broadly \citep{sweeney2002k}. The goal of a re-identification attack is to attach the identity of an individual to a data record without one. That is, re-identification attacks reveal the identity of an individual despite the absence of an ``obvious'' identifier within the data. 

\noindent \textit{Takeaway:} Datasets that have been stripped of explicitly identifying information are not immune to privacy risks. There may still be sufficient information in an ``anonymized'' dataset to reveal the identity of individuals.

\subsubsection{Privacy threats from aggregate statistics and big data computations}
\label{priv_threats_recon_mem}

The previous four harms stem from the possibility of inferring information about an individual beyond what their data explicitly represented. Just as individual data can reveal more than first meets the eye, aggregate statistics -- i.e., statistics computed on multiple individual's data -- can reveal quite a great deal about each individual as well. 

Statistics are numerical summaries of a dataset. Statistics take datasets -- which can contain hundreds, thousands, and even millions of numbers -- and compress them into digestible parcels of information. However, since statistics summarize data, they necessarily contain information about the data. As such, statistics may inadvertently reveal the underlying elements of the dataset. 

The seminal work of \cite{dinur2003revealing} demonstrated that collections of statistics could be analyzed to reverse engineer the input dataset that generated them. For settings where statistics are released because the sharing of individual data is forbidden, such \textit{reconstruction attacks} contradict the purpose of sharing statistics in the first place. The work of Dinur and Nissim, and the subsequent papers that their work inspired, led to what is known as the \textit{Fundamental Law of Information Recovery}: ``too many overly accurate'' statistics necessarily violate the privacy of individuals whose data contributed to the statistics \citep{dwork2019differential}. As such, ``sufficiently noisy'' statistics are required to thwart reconstruction attacks \citep{dwork2017exposed}.\footnote{We discuss how to use differential privacy to construct ``sufficiently noisy'' statistics in Section \ref{3-4_tech_solutions}.} Reconstruction attacks aren't just a theoretical possibility. For example, internal testing by the US Census uncovered that the 2010 census statistics were susceptible to reconstruction attacks, violating the privacy of 46\% of the US population.\footnote{See Amicus Brief of Data Privacy Experts in \textit{Alabama v. US Department of Commerce et al.}, Document 99-1 Filed 04/23/21, Case 3:21-cv-00211-RAH-ECM-KCN.}

Classic summary statistics are not the only information sources vulnerable to reconstruction attacks. Since the parameters of machine learning models are statistics computed on training data, reconstruction attacks can also be levied against machine learning models to reconstruct the training data \citep{balle2022reconstructing}. Machine learning models are also vulnerable to \textit{membership inference attacks},\footnote{Classic statistics are also vulnerable to these sorts of attacks. Within the privacy literature, membership inference attacks are also called \textit{tracing attacks} \citep{dwork2017exposed}.} which aim to determine whether an individual datapoint is in the model's training dataset \citep{shokri2017membership}. The key insight that enables membership inference attacks is that machine learning models -- particularly those that are suffering from overfitting -- tend to be more confident in their predictions on datapoints in their training data \citep{jarin2022miashield}. Consequently, an adversary can query the model with input data, observe the model's confidence on the input, and infer whether the datapoint was likely used to train the model.\footnote{Membership inference attacks, while noxious for individual privacy, can sometimes be used as a tool to audit machine learning algorithms. For instance, in a recent high profile example, \cite{chang2023speak} performed a membership inference attack on ChatGPT and GPT-4 to determine which books these large language models were (likely) trained on.} 

Once individual data records have been reconstructed or inferred to be part of a training dataset, all four of the prior revelatory threats are possible: re-identification attacks can be levied to attach an identity to the data record; inferences can be drawn about the individual from their own record; inferences can be drawn about the individual based on characteristics of similar individuals; and, depending on the dataset in which these records belong to, the social meaning of the dataset may reveal further information about the individual. As we will see in Section \ref{3-4_tech_solutions}, differential privacy can be used to provable thwart reconstruction attacks and membership inference attacks \citep{dwork2017exposed}.

\noindent \textit{Takeaway:} ``Too many overly accurate'' statistics (and other big data computations, such as machine learning models) are vulnerable to reconstruction attacks and membership inference attacks. These attacks allow an adversary to determine what data was used to create a statistic or machine learning model, which can then be used as a starting point to reveal even more information about individuals.

%% file: 3-4_tech_solutions.tex

Technical solutions to data privacy vary based on the specific threat to privacy they seek to mitigate. In general, current technical approaches to computational data privacy seek to protect the underlying data from being ``exposed'' by a computation.%
\footnote{Here, we focus on threats to data privacy that stem from computing statistics on data. In particular, we use the term \textit{statistics} to refer to the output of any computation run on multiple individual's data. Such computations include the calculation of elementary summary statistics (such as the mean, median, and mode) as well as the execution of complicated machine learning training algorithms (that compute the parameters of machine learning models). Additionally, we will refer to data privacy concerns that arise from the computation of statistics as \textit{computational data privacy}. We do so to delineate it from other data privacy concerns, such as the unauthorized sharing of data to third parties, that are important yet distinct from computational concerns.}
There are many ways this exposure occurs. This section explores these different exposures and the current technical solutions that are commonly used to address them.

\subsubsection{Differential Privacy: For Statistics (and Other Big Data Applications)}

In Section \ref{3-3_vulnerabilities}, we discussed how overly accurate statistics and machine learning models are susceptible to reconstruction attacks and membership inference attacks. Fortunately, a mathematical approach to privacy known as \textit{differential privacy} can be used to thwart both of these threats \citep{dwork2017exposed}.

\citet{dwork2006calibrating} introduced the concept of $\epsilon$-indistinguishability, which would subsequently become known as $\epsilon$-differential privacy, to protect statistics from exposing the underlying data they were computed on. Differential privacy provides data subjects with a mathematical promise, via a formal proof, that guarantees ``very little'' will be revealed about them when their data is analyzed \citep{dwork2019differential}. In doing so, differential privacy enables statistical learning from datasets while protecting the privacy of individuals whose data was used to generate the statistics. 

The rationale for constructing a computation that reveals ``very little'' about any individual is as follows: if a statistic reveals ``very little'' about any individual, then any adversary who examines the statistic can only learn ``very little'' about the individuals whose data created the statistic. Hence, data subjects incur ``very little'' privacy loss when their data are analyzed. This notion of ``very little'' is made mathematically precise with the introduction of a privacy parameter $\epsilon$ (read as ``epsilon''). As such, smaller values of $\epsilon$ correspond to smaller privacy losses, and larger values of $\epsilon$ correspond to larger privacy losses.

In order to achieve this mathematical promise, differentially private algorithms introduce noise when computing statistics.\footnote{The presence of any noise in a statistical calculation does not automatically make the statistic differentially private. The noise that differential privacy requires is very particular -- it must be carefully calibrated and come from appropriate noise distributions \citep{dwork2006calibrating}.} That is, differentially private algorithms release ``noisy statistics'' instead of true statistics in order limit the statistic's ability to expose individual data. 
The amount of noise introduced is related to the privacy parameter $\epsilon$. While smaller values of $\epsilon$ yield more privacy, they introduce more noise into the statistic, yielding less accurate results. Alternatively, larger values of $\epsilon$ require the introduction of less of noise, which provides more accurate results at the expense of weaker privacy protections. For this reason, differential privacy induces a \textit{privacy-accuracy} tradeoff, which is controlled by the parameter $\epsilon$. The value $\epsilon$ is a tunable by the algorithm developer, which allows for different tolerances for the privacy-accuracy tradeoff to be achieved in practice. Since different datasets may have different sensitivities about them, and since different statistics may require different levels of accuracy to be useful, there is no universal consensus on how to set $\epsilon$ \citep{dwork2019differential}. However, several different methods exist to identify appropriate values of $\epsilon$ based on context-specific goals  \citep{hsu2014differential, kohli2018epsilon, dwork2019differential, kohli2023privacy}.

There are many different algorithms that one can use to achieve differential privacy, such as the Laplace mechanism \citep{dwork2006calibrating}, Gaussian mechanism \citep{dwork2006our}, multiplicative weights mechanism \citep{hardt2010multiplicative}, and general mechanisms for arbitrary statistical analyses \citep{mohan2012gupt, kohli2023differential}. Since the parameters of a machine learning model are statistics computed from training data, differentially private algorithms can be leveraged to produce differentially private machine learning models \citep{ji2014differential, abadi2016deep, papernot2018scalable, ha2019differential}. We note, however, that differential privacy is often incorrectly described as a process by which noise is added to the data to protect privacy. While some differentially private algorithms may perturb individual data,\footnote{A variant of differential privacy, known as \textit{local differential privacy}, can be used to perturb user information. We will elaborate more on this in Section \ref{local_dp}.} other algorithms do not. In fact, many algorithms that satisfy differential privacy perturb the output of a statistic while leaving the original data intact.

Differential privacy is accompanied with a whole host of formal guarantees that make it useful for privacy engineering. First, differential privacy makes it possible to keep track of the cumulative privacy loss accrued over multiple analyses. For instance, suppose an analyst runs an $\epsilon_1$-differentially private algorithm on a dataset and then runs an $\epsilon_2$-differentially private algorithm on the same data. In the first analysis, data subjects have incurred $\epsilon_1$ privacy loss, whereas in the second analysis, data subjects have incurred $\epsilon_2$ privacy loss. So in total, data subjects have incurred is $\epsilon_1+\epsilon_2$ privacy loss. This \textit{composition result} informs us that, when multiple differentially private analyses are run on a dataset, the privacy loss is cumulative. For this reason, differential privacy enables rigorous accounting of privacy loss. 

The composition result tells us what happens when multiple private analyses are performed on the same dataset. Another property of differential privacy, known as the \textit{post-processing result}, tells us what happens when someone analyzes a differentially private statistic: without access to the original data, data subjects do not incur additional privacy loss when the output of differentially private computation is analysed. That is, differential privacy fully inoculates a statistic in such a manner to ensure that no matter what anyone else uses this statistic for, data subjects will incur no further privacy loss. Fascinatingly, this result holds \textit{regardless} of (1) the auxiliary information an adversary may have, (2) the specific attack strategy they employ, and (3) the computing power and technology they have -- now or in the future. For this reason, differential privacy is often said to be \textit{future proof and adversary agnostic} \citep{dwork2019differential}.

\subsubsection{Differential Privacy: For Synthetic Data}

While differential privacy was introduced as a means to compute privacy-preserving statistics, techniques from differential privacy have been used to generate privacy-preserving synthetic datasets. Conceptually, a synthetic dataset is a ``fake dataset'' that can be used in place of a real dataset for computations. To ensure that synthetic datasets are representative of the real dataset they seek to replace, synthetic datasets are constructed to replicate numerous statistical properties of the real dataset. 

By themselves, synthetic datasets have no formal privacy guarantee. Since statistics can expose underlying elements of a dataset, synthetic data built from these statistics can similarly expose elements of the real dataset \citep{giomi2022unified}.\footnote{For a gentle and conceptual introduction to synthetic data, see \cite{ulman2022synthetic}.} However, when a synthetic dataset is constructed using techniques from differential privacy, the dataset inherits all of the protections that differential privacy has to offer. Most relevant for differentially private synthetic data is the post-processing guarantee: any use of a synthetic dataset will not incur additional privacy loss for data subjects. This is intuitively appealing for many data analysis tasks, as differentially private synthetic data would enable anyone to analyze the synthetic data without compromising the privacy of the individuals in the real dataset. 

In their seminal paper, \cite{blum2008learning} showed that it is possible to construct a differentially private synthetic dataset that preserves a large number of statistical characteristics of the original data. However, their method is computationally expensive, making it impractical to employ in practice (except on small datasets). Currently, the creation of differentially private synthetic data is an active and promising area of research \citep{zhang2021privsyn, aydore2021differentially, tantipongpipat2021differentially, bun2023continual}.

\subsubsection{Homomorphic Encryption}

Differential privacy aims to protect a statistic from exposing elements of the underlying dataset. In this sense, differential privacy seeks to protect the data \textit{post-computation} by inoculating the statistic with carefully calibrated noise. In order to protect data from being exposed \textit{during the computation process}, other privacy paradigms are required. One such paradigm is \textit{homomorphic encryption}.\footnote{For a more technical introduction to homomorphic encryption, see \cite{halevi2017homomorphic}.}

The concept of homomorphic encryption was introduced by  \cite{rivest1978data}. The goal of homomorphic encryption is to design an encryption protocol in such a manner that permits computations over  encrypted data without the need to decrypt it. Homomorphic encryption derives its name from these two desiderata: \textit{homo} (meaning same) and \textit{morphic} (meaning structure). By design, homomorphic encryption ensures that the mathematical structure of computations over encrypted data is the same as in the unencrypted state. 

The first fully homomorphic scheme -- a scheme that allows for arbitrary computations -- was introduced by \cite{gentry2009fully}. This scheme, however, was computationally inefficient to implement. Since then, other homomorphic encryption schemes have been introduced that are more efficient, but restrict the types of computations that can be performed \citep{evans2017pragmatic}. Despite these restrictions, schemes exist that allow for addition, subtraction, and multiplication, which can be applied in many practical settings involving statistics and machine learning. One prominent example was the 2016 advent of \textit{CryptoNets}, which allows for the training of neural networks on homomorphically encrypted data \citep{gilad2016cryptonets}. 

While both homomorphic encryption and differential privacy operate in the statistical setting, they address different privacy concerns. Differential privacy is concerned with attacks on statistics that can be used to reveal information about individuals in the underlying dataset post-computation. Homomorphic encryption, on the other hand, is concerned with attacks that can occur when data is in an unencrypted form during computation. Both are concerned about protecting the underlying data -- however, they provide protection against different threats that arise at different points in the computation process.

\subsubsection{Secure Multiparty Computation}

So far we have considered computational settings where data resides in a central location. In some situations, however, data is not able to housed in a single location. This can occur for a myriad of reasons, such as a lack of trust between multiple data holders, as well as legal restrictions that outright forbid the sharing of data. 

In situations such as these, \textit{decentralized computing} -- also known as \textit{distributed computing} -- can be used to calculate statistics from data without the need to store the data in a central location. While the lack of a single central repository of data can alleviate data privacy concerns surrounding the storage of information, decentralization of data alone does not yield any formal privacy guarantees. However, when combined with other privacy paradigms, formal privacy guarantees can be established. 

One such paradigm is \textit{secure multiparty computation}. Secure multiparty computation, often referred to as multiparty computation (MPC), was introduced by  \cite{yao1982protocols} to enable the computation of statistics that rely on multiple parties' data, such that each party learns nothing more about another party's data other than what is revealed by the statistic. MPC is established through protocols that specify the way in which parties act so as to properly compute information in such a manner that, when aggregated together, the result is correct and does not expose any party's input data. 

The first protocol was Yao’s garbled circuit protocol (GPC), which can be used to compute an arbitrary discrete function (provided that this function can be represented as a fixed-size circuit) \citep{evans2017pragmatic}. To date, Yao's GCP remains a foundational building-block for many MPC applications. MPC applications may also use other protocols, such as those involving secret sharing and homomorphic encryption. 

Recently, researchers partnered with two Dutch banks to perform privacy-preserving money laundering detection using MPC \citep{sangers2019secure, van2021privacy}. Since data sharing between financial institutions raises privacy concerns, MPC enabled the banks to collaboratively compute over each other's data to detect money laundering without the need to ever share consumer data.

\subsubsection{Local Differential Privacy}
\label{local_dp}

In similar spirit to decentralization, a variant of differential privacy, known as \textit{local differential privacy} (LDP), was introduced to compute privacy-preserving statistics in settings where personal data needs to remain private even from the data collector. In our prior discussion of differential privacy, a central authority was tasked with computing noisy statistics from centrally located data. In this case, noise is introduced into the statistic \textit{after} the central authority receives the data. As such, this classic setting is often called the \textit{central} differential privacy setting. 

In the LDP setting, a central authority is still tasked with computing noisy statistics. However, in this setting  noise is introduced into information \textit{before} the central authority receives it. This is appealing at an intuitive level, as the central authority never receives real personal information -- they only receive noisy personal information. 

To elucidate discussion, consider a financial institution that wants to understand the reliability of their mobile banking app. The financial institution will assess this by examining the total number of app crashes across their entire user base. Since the company is interested in the total number of crashes, the company could request a noisy count of crashes from each subscriber's phone. Summing up the noisy individual estimates would provide a noisy estimate of the total number of app crashes across the company's user base without ever requiring the transfer of crash logs that could contain sensitive mobile phone metadata. As was the case with (central) differential privacy, a wide variety of mechanisms exist to achieve local differential privacy \citep{erlingsson2014rappor, kairouz2016extremal, bebensee2019local}

\subsubsection{Federated Learning}

Similar in spirit to MPC is \textit{federated learning} (FL), which enables the decentralized development of machine learning models \citep{mcmahan2017communication}. Following on the theme of decentralization, the idea of FL is to develop a single machine learning model based on $n$ individuals data without the need to consolidate individual data prior to the model's development.

At a high level, FL enables the local training of models in such a manner that the local model updates can be combined to a global model update. Federated learning presents an opportunity to train machine learning models from individual data without the need for personal data transfer. Instead, local model updates are transferred for global model development. 

However, as noted above, decentralization alone does not provide a rigorous privacy guarantee. Absent any additional requirements, local model updates can leak information about the local data. When coupled with mathematically rigorous techniques from MPC and LDP, FL can now be done in a manner with strong privacy guarantees. For example, \cite{truex2019hybrid} showed that a wide array of machine learning models (such as decision trees, neural networks, and support vector machines) could be trained in a federated manner in conjunction with techniques from MPC and LDP. 



\subsubsection{Technical Privacy Solutions for LMICs: What Research is Needed?}


Techniques from differential privacy (in both the central and local settings), homomorphic encryption, secure multiparty computation, and federated learning can be used to address different privacy concerns related to the exposure of sensitive information. Recently, promising research has emerged that combines multiple techniques from these privacy paradigms to afford individuals stronger levels of privacy \citep{wagh2021dp}. This remains and active and exciting area of research within computer science and computer security.

\paragraph{\textit{`Technical Solution' Research Questions}} However, the vast majority of these innovations and advances have been developed for use cases and markets in developed economies -- typically by companies operating in those markets or by researchers receiving support from companies or funding agencies from those markets. By contrast, very little -- if any -- research  has been done on privacy-preserving big data applications in emerging markets. For billions of users, this leaves many important questions unanswered: 

\begin{enumerate}
    \item Which, if any, of the existing privacy-enhancing technologies could be used ``off the shelf'' in emerging market applications?  Are there particular settings where minimal tinkering is required? 
    \item How can these approaches be used in conjunction with one another to mitigate a wider family of privacy harms? Would this more adequately address the key privacy concerns in emerging markets?
    \item Are there particular settings where existing technical solutions fail to address the underlying privacy concerns that are most important in that setting? 
    \item What new technical innovations can best protect the data privacy concerns that are unique to emerging markets in general, or specific local contexts?
\end{enumerate}


%% file: 3-5_nontech_solutions.tex
In addition to understanding the potential and limitations of technical solutions to data privacy in emerging markets, \textit{non-technical} approaches -- such as policies and practices -- can also be very effective at proactively addressing privacy concerns. 
For instance, in the US context, government organizations are frequently bound by written laws and policies that dictate appropriate uses of data \citep{bushkin1976privacy, officer2009privacy}. In Europe, privacy restrictions tend to be more stringent. For instance, the General Data Protection Regulation (GDPR) places legal restrictions on the collection, storage, analysis, and sharing of personal data. And even when the use of personal data is admissible, the GDPR requires compliance with seven data protection and accountability principles. Failure to abide by the GDPR can result in a maximum fine of 4\% of global revenue or \euro20 million, whichever is larger.\footnote{See \url{https://gdpr.eu/what-is-gdpr/}.} These stringent fines aim to incentivize compliance with the GDPR, and hence incentivize proactive data privacy practices.

In the private sector, corporations also frequently adopt data use guidelines that enable employees and products to better attend to privacy \citep{bamberger2015privacy}. While these policies vary across organizations, sector-specific privacy and data protection laws provide baseline requirements on how and when information may be collected, stored, protected, accessed, used, and deleted. Common practices include data minimization, use limitation, sharing restrictions and requirements, and  de-identification \citep{sedenberg2015public}. While these policies lay out the rules of the road, and while these practices help organizations stay in the correct lane, additional accountability mechanisms, such as access controls and audit logging, can also be used to detect deviations from these policies and practices \citep{bowman2015architecture}.



However, to ensure that policies and practices are well-calibrated to the setting at hand, these polices and practices must build off of local conceptions of privacy. As noted earlier in Section \ref{3-2_current_priv_defs}, several prominent researchers have argued that the ``Silicon Valley paradigm'' of data privacy is inappropriate in emerging market settings \citep{abebe2021narratives}, meaning that many data sharing practices that are common in the US may be viewed as privacy violations in the Global South. 

\paragraph{\textit{`Non-Technical Solution' Research Questions}}Since privacy is a social concept, actions that violate norms of expected uses of data are often viewed as privacy violations by data subjects \citep{nissenbaum2004privacy}. Consequently, there is a need for qualitative research to understand \textit{what} particular uses of data are viewed as appropriate by a community within a given setting, and what uses of data are viewed as inappropriate as well.  In order to understand potential mismatches in current data privacy policies and practices, several questions require deeper investigation.

\begin{enumerate}
    \item How does current or proposed regulation address potential privacy harms and consumers' expectations of privacy? How does it fall short? To what extent is regulation enforced, and to what extent does it more broadly affect product innovation?
    \item How can financial service providers best  engage proactively with customers -- by providing information, requesting consent, etc. -- to better align customer expectations with current or future practice?
    \item To what extent do existing practices align with consumer's expectations of privacy? Where do these practices miss the mark? Is it with respect to information collection, access, use, sharing, or deletion? And what new practices can be implemented to build and foster consumers' trust that their local notions of privacy are being respected?
\end{enumerate}

%% file: 3-6_privacy_values_tradeoff.tex
Strong privacy guarantees do not come for free. For example, differentially private machine learning algorithms frequently induce a tradeoff between privacy and accuracy: the more private the system, the less accurate the algorithm. When those algorithms are used to drive key decisions -- such as credit scoring and risk management -- the loss in accuracy can have a direct consequence on the bottom line (see also the discussion in Section~\ref{3-4_tech_solutions}).

In some settings, quantitative analysis can mathematically demonstrate the exact nature of the privacy-accuracy tradeoff for a given context \citep{kohli2023privacy}. For instance, it may be possible to determine that an X\% increase in privacy would translate to a $Y$\% decrease in accuracy (and thus a $Z$\% loss in profits). In the context of digital credit, for example, it may be possible to construct a credit score based on a private version of the borrower's personal data -- but that credit score would be less accurate than a version constructed from the raw data, and as such, might not be as profitable for the financial institution.
In other cases, lower and upper bounds can be derived to describe the range of the privacy-accuracy tradeoff. But this type of rigorous analysis is rarely documented even in places like the US; to our knowledge, no such case studies exist from emerging markets.\footnote{One exception is our recent work in \cite{kohli2023privacy}, available \href{https://arxiv.org/pdf/2306.09471.pdf}{here}, that characterizes the privacy-accuracy tradeoff of using differentially private mobility matrices in humanitarian response.}

The accuracy of statistics and machine learning models can be tied to many different objectives depending on what they are used for. When statistics are used for business decisions, accurate statistics promote profitability. Other times, statistics may be used to inform public health interventions and humanitarian response. In these settings, accurate statistics promote social welfare. Alternatively, US Census statistics are used by the government for a whole host of tasks, such as providing election information in minority languages and allocation educational funds \citep{pujol2020fair}. In this setting, accurate statistics promote dignity, representation, and fairness in governance. 



Improvements in privacy often involve a wider range of tradeoffs beyond the privacy-accuracy tradeoff that is salient with differential privacy \citep{kohli2021leveraging}. For instance, recent machine learning research has documented a frequent tradeoff between privacy and \textit{fairness}: in settings where fairness criteria are not explicitly built into the learning process, differentially private algorithms can exacerbate inequities \citep{bagdasaryan2019differential}. In other settings, there may exist similarly pointed tradeoffs between privacy and other social or corporate objectives such as transparency (more transparent decision rules may more visibility into user data), social welfare (maximizing social welfare may require accessing private information), and competition (monopolistic behavior could be inhibited by data sharing requirements), to name just a few examples.


To complicate matters further, tradeoffs are often multidimensional \citep{rolf_balancing_2020}. For example, a lending institution may want to preserve privacy and ensure that their lending algorithm does not discriminate on protected attributes, but they may also want to ensure that the algorithm still promotes profitability. When multiple objectives are considered, striking a balance between effective privacy protection and other objectives is a non-trivial and highly contextual task. 

\paragraph{\textit{``Tradeoffs'' Research Questions}}If and how these privacy tradeoffs relate to emerging markets is very poorly understood, leaving many basic questions unanswered:

\begin{enumerate}
    \item What are the theoretical and empirical tradeoffs between privacy and accuracy in real-world use cases (such as with alternative credit scores)? How do the reductions in accuracy translate to other key corporate objectives (like risk and profits)?
    \item In addition to accuracy, does data privacy induce tradeoffs with other key performance indicators? How can those tradeoffs best be documented and quantified?
    \item To what extent do different notions of privacy -- and in particular those that are most relevant in emerging markets -- make the tradeoff between privacy and accuracy more or less stark?
    \item Are financial service providers willing to make small sacrifices in order to protect consumer privacy? If so, under what circumstances?
\end{enumerate}

%% file: 4_conclusion.tex
To enable continued use and innovation of personal big data in financial services and related emerging market applications, we believe it is critical to launch a coordinated research agenda to develop a more complete understanding of the problem and solution space of big data privacy in low- and middle-income countries. Too little is known about how to most effectively protect consumers in these markets, or even of what the biggest threats to consumer privacy might be. This lack of knowledge creates systemic risk that threatens to undermine service innovation and adoption. We believe that proactively identifying and addressing potential threats and opportunities \textit{now} --- rather than in reaction to a data privacy scandal --- is the best way to ensure the health and welfare of consumers as well as the fintech industry.

To seed this research agenda, this paper provides an overview of modern approaches to data privacy, and identifies five major themes for understanding how best to adapt and improve those methods for use in emerging markets. These themes include:

\begin{enumerate}
    \item A landscape analysis of key use cases and current privacy protections in financial services and related emerging market applications, with a focus on identifying key vulnerabilities, current best practices, and gaps in knowledge and technology.
    \item Defining ``data privacy'' in emerging markets, with a focus on understanding if and how the concerns coming from low-income countries differ from those found in wealthy nations.
    \item Expanding and adapting technical solutions to data privacy for use in emerging markets. Key technologies include differential privacy, homomorphic encryption, secure multiparty computation, and federated learning. To date, however, these technologies have been almost exclusively developed and deployed in high-income countries.
    \item Developing and improving non-technical approaches to data privacy (such as policies and practices) that complement technical solutions.
    \item Understanding the tradeoffs incurred by financial institutions, fintechs, and individuals when privacy-enhancing solutions are deployed in practice. 
\end{enumerate}

The increasing ubiquity of digital technology in emerging markets has the ability to revolutionize fintech and financial service provision in low- and middle-income countries. However, inadequate consideration for data privacy has the potential to undermine this innovation. This need not be the case. Appropriately leveraged, privacy-enhancing technologies, policies, and practices have the ability to simultaneously enable innovative uses of big data while also  providing robust and meaningful privacy protections for personal data. We believe this research agenda can help catalyze a more responsible and consumer-oriented approach to big data innovation in financial services.